%Paper: gr-qc/9506065
%From: ESPOSITO@napoli.infn.it
%Date: Wed, 28 Jun 1995 8:21:43 +0200 (CET-DST)

\magnification \magstep1
\raggedbottom
\openup 4\jot
\voffset6truemm
\headline={\ifnum\pageno=1\hfill\else
\hfill{\it Local Supersymmetry in One-Loop Quantum Cosmology}
\hfill \fi}
\centerline {\bf LOCAL SUPERSYMMETRY IN ONE-LOOP}
\centerline {\bf QUANTUM COSMOLOGY}
\vskip 1cm
\centerline {\bf Giampiero Esposito}
\vskip 1cm
\centerline {\it Istituto Nazionale di Fisica Nucleare}
\centerline {\it Mostra d'Oltremare Padiglione 20, 80125 Napoli, Italy;}
\centerline {\it Dipartimento di Scienze Fisiche}
\centerline {\it Mostra d'Oltremare Padiglione 19, 80125 Napoli, Italy.}
\vskip 1cm
\noindent
{\bf Abstract.} The contribution of physical degrees of freedom to
the one-loop amplitudes of Euclidean supergravity is here evaluated
in the case of flat Euclidean backgrounds bounded by a three-sphere,
recently considered in perturbative quantum cosmology.
In Euclidean supergravity, the spin-${3\over 2}$
potential has the pair of independent spatial components
$\Bigr(\psi_{i}^{A}, \; {\widetilde \psi}_{i}^{A'}\Bigr)$.
Massless gravitinos are here
subject to the following local boundary conditions
on $S^3$: $\sqrt{2} \; {_{e}n_{A}^{\; \; A'}}
\psi_{i}^{A}=\pm {\widetilde \psi}_{i}^{A'}$,
where ${_{e}n_{A}^{\; \; A'}}$ is the Euclidean
normal to the three-sphere boundary.
The physical degrees of freedom (denoted by PDF)
are picked out imposing the supersymmetry
constraints and choosing the gauge condition
$e_{AA'}^{\; \; \; \; \; i}\psi_{i}^{A}=0$,
$e_{AA'}^{\; \; \; \; \; i}{\widetilde \psi}_{i}^{A'}=0$.
These local boundary conditions are then
found to imply the eigenvalue condition
${\Bigr[J_{n+2}(E)\Bigr]}^{2}-{\Bigr[J_{n+3}(E)\Bigr]}^{2}=0,
\; \forall n \geq 0$, with degeneracy
$(n+4)(n+1)$. One can thus apply again a
zeta-function technique previously
used for massless spin-${1\over 2}$ fields.
The PDF contribution to the full $\zeta(0)$ value is found to be
$=-{289\over 360}$. Remarkably, for the
massless gravitino field the PDF method and
local boundary conditions lead to
a result for $\zeta(0)$ which is equal to the PDF value one obtains
using spectral boundary conditions on $S^{3}$.
\vskip 13cm
\leftline {PACS numbers: 03.70.+k, 04.60.+n, 98.80.Dr}
\vskip 100cm
\leftline {\bf 1. Introduction}
\vskip 1cm
\noindent
The problem of one-loop finiteness of supergravity theories
in the presence of boundaries is still receiving careful
consideration in the current literature.$^{1-10}$ As emphasized
in Refs. 9,11-12, one can perform
one-loop calculations paying attention
to: (1) S-matrix elements; (2) topological invariants;
(3) presence of boundaries. For example, in the case of pure
gravity with vanishing cosmological constant: $\Lambda=0$, it is
known that one-loop on-shell S-matrix elements are finite.
This property is shared by $N=1$ supergravity when
$\Lambda=0$, and in that theory two-loop on-shell finiteness
also holds. However, when $\Lambda \not = 0$, both pure gravity
and $N=1$ supergravity are no longer one-loop finite in the
sense (1) and (2), because the non-vanishing on-shell one-loop
counterterm$^{11}$ is given by
$$
S_{(1)}={1\over \epsilon}\biggr[A\chi-{2BG\Lambda S \over
3\pi}\biggr]
\; \; \; \; .
\eqno (1.1)
$$
In equation (1.1), $\epsilon=n-4$ is the
dimensional-regularization parameter, $\chi$ is the Euler
number, $S$ is the classical on-shell action, and one finds
:$^{9,11}$ $A={106\over 45}, B=-{87\over 10}$ for pure gravity,
and $A={41\over 24}, B=-{77\over 12}$ for $N=1$ supergravity.
Thus, $B \not =0$ is responsible for lack of S-matrix one-loop
finiteness, and $A \not =0$ does not yield topological one-loop
finiteness.

If any theory of quantum gravity can be studied from a
perturbative point of view, boundary effects play a key role
in understanding whether it has interesting and useful
finiteness properties. It is therefore necessary to analyze
in detail the structure of the one-loop boundary counterterms
for fields of various spins. This problem has been recently
studied within the framework of one-loop quantum cosmology,
where the boundary is usually taken to be a three-sphere,
and the background is flat Euclidean space or a de Sitter
four-sphere or a more general curved four-geometry.$^{2-10}$

Our paper describes one-loop properties of spin-${3\over 2}$
fields to present a calculation which was previously studied
in research books$^{5,9}$ but not in physics journals (see,
however, remarks at the end of Ref. 10). In the Euclidean-time
regime, the spin-${3\over 2}$ field is represented, using
two-component spinor notation, by a pair of independent
spinor-valued one-forms $\Bigr(\psi_{\mu}^{A},
{\widetilde \psi}_{\mu}^{A'}\Bigr)$ with spatial components
$\Bigr(\psi_{i}^{A},
{\widetilde \psi}_{i}^{A'}\Bigr)$.$^{4,9}$
After imposing the gauge conditions (hereafter
$e_{AA'}^{\; \; \; \; \; \mu}$ is the tetrad)
$$
e_{AA'}^{\; \; \; \; \; i} \; \psi_{i}^{A}=0
\; \; \; \; , \; \; \; \;
e_{AA'}^{\; \; \; \; \; i}
{\widetilde \psi}_{i}^{A'}=0
\; \; \; \; .
\eqno (1.2)
$$
and the linearized supersymmetry constraints, the expansion
of $\Bigr(\psi_{i}^{A},{\widetilde \psi}_{i}^{A'}\Bigr)$
on a family of three-spheres centred on the origin takes the
form$^{4,9}$
$$
\psi_{i}^{A}={\tau^{-{3\over 2}}\over 2\pi}
\sum_{n=0}^{\infty}\sum_{p,q=1}^{(n+1)(n+4)}
\alpha_{n}^{pq}
\biggr[m_{np}(\tau)\beta^{nqABB'}
+{\widetilde r}_{np}(\tau){\overline \mu}^{nqABB'}
\biggr]e_{BB'i}
\; \; ,
\eqno (1.3)
$$
$$
{\widetilde \psi}_{i}^{A'}={\tau^{-{3\over 2}}\over 2\pi}
\sum_{n=0}^{\infty}\sum_{p,q=1}^{(n+1)(n+4)}
\alpha_{n}^{pq}
\biggr[{\widetilde m}_{np}(\tau)
{\overline \beta}^{nqBA'B'}
+r_{np}(\tau)\mu^{nqBA'B'}\biggr]e_{BB'i}
\; \; .
\eqno (1.4)
$$
With our notation, $\tau$ is the radial distance from the
origin in flat Euclidean four-space, the matrix
$\alpha_{n}^{pq}$ is block-diagonal in the indices $pq$,
with blocks $\pmatrix {1&1\cr 1&-1\cr}$. Note also that the
modes ${\widetilde m}_{np}(\tau), {\widetilde r}_{np}(\tau)$
are not the complex conjugates of $m_{np}(\tau),r_{np}(\tau)$
respectively. Moreover, one has$^{4,9}$
$\beta^{nqABB'}=-\rho^{nq(ABC)}n_{C}^{\; \; B'} \; ,
\mu^{nqBA'B'}=-\sigma^{nq(A'B'C')}n_{\; \; C'}^{B}$
where the harmonics $\rho^{nq(ABC)}$ and $\sigma^{nq(A'B'C')}$
are symmetric in their three spinor indices and have positive
eigenvalues ${1\over 2}\Bigr(n+{5\over 2}\Bigr)$ of the
intrinsic three-dimensional Dirac operator on $S^{3}$,$^{4}$
and $n^{CB'}$ is the Lorentzian normal to $S^{3}$.$^{3,9}$

Sec. 2 studies locally supersymmetric boundary conditions
on $S^3$ for the spin-${3\over 2}$ potential, and the equation
obeyed by the eigenvalues by virtue of these boundary conditions
is derived. Sec. 3 uses zeta-function regularization and
obtains the contribution of physical degrees of freedom
(hereafter referred to as PDF) to the
full $\zeta(0)$ value. Concluding remarks and open research
problems are presented in Sec. 4.
\vskip 1cm
\leftline {\bf 2. Local Boundary Conditions for the Spin-${3\over 2}$
Potential}
\vskip 1cm
\noindent
In Euclidean supergravity, the mathematical
description of the gravitino leads to the introduction of
the independent spinor-valued one-forms
$\Bigr(\psi_{\mu}^{A}, \; {\widetilde \psi}_{\mu}^{A'}\Bigr)$ with spatial
components $\Bigr(\psi_{i}^{A}, \; {\widetilde \psi}_{i}^{A'}\Bigr)$.
We are here interested in a generalization to simple
supergravity of the calculations in Ref. 3 for the spin-${1\over 2}$
field. Thus, we consider a flat Euclidean background,
requiring on the bounding $S^3$ that
$$
\sqrt{2} \; {_{e}n_{A}^{\; \; A'}} \psi_{i}^{A}=\epsilon \;
{\widetilde \psi}_{i}^{A'}
\; \; \; \; ,
\eqno (2.1)
$$
where $\epsilon = \pm 1$.
The consideration of (2.1) is suggested by the work in
Ref. 1, where it is shown that the spatial tetrad
$e_{\; \; \; \; \; i}^{AA'}$ and the projection
$\Bigr(\pm {\widetilde \psi}_{i}^{A'}-\sqrt{2}\;
{_{e}n_{A}^{\; \; A'}}\psi_{i}^{A}\Bigr)$
transform into each other under half of the local supersymmetry
transformations at the boundary,
and that after adding a suitable boundary term,
the supergravity action is invariant under these local supersymmetry
transformations.$^{3,9}$

Indeed, from Sec. 1 we already know that, imposing the supersymmetry
constraints and choosing the gauge condition (1.2),
the spin-${3\over 2}$ potential finally assumes the form (1.3)-(1.4).
It is therefore useful to derive identities relating barred to unbarred
harmonics, generalizing the technique in Ref. 13.
This is achieved by using the relations
$$
\int d\mu \; \rho_{ABC}^{np}n^{AA'}n^{BB'}n^{CC'}
{\overline \rho}_{A'B'C'}^{mq}=\delta^{nm}H_{n}^{pq}
\; \; \; \; ,
\eqno (2.2)
$$
$$
\int d\mu \; \rho_{ABC}^{np}\epsilon^{AD}\epsilon^{BE}\epsilon^{CF}
\rho_{DEF}^{mq}=\delta^{nm}A_{n}^{pq}
\; \; \; \; ,
\eqno (2.3)
$$
and the expansion of the totally symmetric field strength
$$
\phi_{ABC}(x)=\sum_{n=0}^{\infty}\sum_{p=1}^{(n+1)(n+4)}
\Bigr({\widehat a}_{np}\rho_{ABC}^{np}(x)
+{\widehat b}_{np}
{\overline \sigma}_{ABC}^{np}(x)\Bigr) \; \; \; \; .
\eqno (2.4)
$$
Thus, we can express the
${\widehat a}_{np}$ coefficients in two equivalent ways using
(2.4), and (2.2) or (2.3). The equality of the two
resulting formulae leads to
$$
n^{AA'}n^{BB'}n^{CC'}\sum_{q=1}^{(n+1)(n+4)}
{\overline \rho}_{A'B'C'}^{nq}
{\left(H_{n}^{-1}\right)}^{qp}=\epsilon^{AD}\epsilon^{BE}
\epsilon^{CF}\sum_{q=1}^{(n+1)(n+4)}\rho_{DEF}^{nq}
{\left(A_{n}^{-1}\right)}^{qp}
\; ,
\eqno (2.5)
$$
which is finally cast in the form
$$
{\overline \rho}_{D'E'F'}^{np}=-8n_{\; \; D'}^{D}n_{\; \; E'}^{E}
n_{\; \; F'}^{F}\sum_{q=1}^{(n+1)(n+4)}\rho_{DEF}^{nq}
{\left(A_{n}^{-1}H_{n}\right)}^{qp}
\; \; \; \; .
\eqno (2.6)
$$
In a similar way, we obtain
$$
{\overline \sigma}_{DEF}^{np}=-8n_{D}^{\; \; D'}n_{E}^{\; \; E'}
n_{F}^{\; \; F'}\sum_{q=1}^{(n+1)(n+4)}\sigma_{D'E'F'}^{nq}
{\left(A_{n}^{-1}H_{n}\right)}^{qp} \; \; \; \; .
\eqno (2.7)
$$
The form of the matrices $A_{n}^{pq}$ and $H_{n}^{pq}$ is obtained taking the
complex conjugate of (2.6), and then inserting the form of
$\rho_{DEF}^{np}$ so obtained into the right-hand side of (2.6).
This yields the consistency condition
$$
A_{n}^{-1}H_{n}A_{n}^{-1}H_{n}=-{1\over 8}1_{n}
\; \; \; \; ,
\eqno (2.8)
$$
which is solved by $A_{n}^{-1}={1\over 2 \sqrt{2}}
\pmatrix{
0&-1 \cr
1&0 \cr }$, $H_{n}= \pmatrix{ 1&0 \cr 0&1 \cr}$,
so that $A_{n}=2\sqrt{2} \pmatrix{ 0&1 \cr -1&0 \cr}$. We can now remark that
(1.3)-(1.4) and (2.1) imply
$$
-i\sqrt{2}\sum_{p,q=1}^{(n+1)(n+4)}
\alpha_{n}^{pq}m_{np}^{(\beta)}(a)\rho^{nqABD}
n_{A}^{\; \; A'}n_{D}^{\; \; B'}=\epsilon
\sum_{p,q=1}^{(n+1)(n+4)}
\alpha_{n}^{pq}{\widetilde m}_{np}^{(\beta)}(a)
{\overline \rho}^{nqA'B'D'}n_{\; \; D'}^{B}
\; ,
\eqno (2.9)
$$
$$
-i\sqrt{2}\sum_{p,q=1}^{(n+1)(n+4)}
\alpha_{n}^{pq}{\widetilde r}_{np}^{(\mu)}(a)
{\overline \sigma}^{nqABD}n_{A}^{\; \; A'}n_{D}^{\; \; B'}
=\epsilon \sum_{p,q=1}^{(n+1)(n+4)}
\alpha_{n}^{pq}r_{np}^{(\mu)}(a)\sigma^{nqA'B'D'}
n_{\; \; D'}^{B} \; \; \; \; .
\eqno (2.10)
$$
This is why Eqs. (2.6)-(2.7),
(2.9)-(2.10) and the formulae for
$A_{n}^{-1}H_{n}$ lead to the boundary conditions
$$ \eqalignno{
i\sum_{pq} {\pmatrix {1&1\cr 1&-1\cr}}^{pq}
m_{np}^{(\beta)}(a)\rho^{nqABC}&=\epsilon
\sum_{pq}{\pmatrix {1&1\cr 1&-1\cr}}^{pq}
{\widetilde m}_{np}^{(\beta)}(a) \cdot \cr
&\cdot \sum_{d}\rho^{ndABC}
{\pmatrix {0&-1\cr 1&0\cr}}^{dq}
\; \; ,
&(2.11)\cr}
$$
$$ \eqalignno{
-\epsilon \sum_{pq}{\pmatrix {1&1\cr 1&-1\cr}}^{pq}
r_{np}^{(\mu)}(a)\sigma^{nqA'B'C'}
&=i\sum_{pq}{\pmatrix {1&1\cr 1&-1\cr}}^{pq}
{\widetilde r}_{np}^{(\mu)}(a) \cdot \cr
&\cdot \sum_{d}\sigma^{ndA'B'C'}
{\pmatrix {0&-1\cr 1&0\cr}}^{dq}
\; \; .
&(2.12)\cr}
$$
Since the $\rho$- and $\sigma$-harmonics on the bounding
three-sphere of radius $a$ are
linearly independent, the typical case of
the indices $p,q=1,2$ yields$^{3,9}$
$$
im_{n1}^{(\beta)}(a)=\epsilon \; {\widetilde m}_{n2}^{(\beta)}(a)
\; \; \; \; ,
\eqno (2.13)
$$
$$
-im_{n2}^{(\beta)}(a)=\epsilon \; {\widetilde m}_{n1}^{(\beta)}(a)
\; \; \; \; ,
\eqno (2.14)
$$
$$
i{\widetilde r}_{n1}^{(\mu)}(a)=\epsilon \; r_{n2}^{(\mu)}(a)
\; \; \; \; ,
\eqno (2.15)
$$
$$
-i{\widetilde r}_{n2}^{(\mu)}(a)=\epsilon \; r_{n1}^{(\mu)}(a)
\; \; \; \; .
\eqno (2.16)
$$
If we now set $\kappa_{n} \equiv n+{5\over 2}$ and define,
$\forall n \geq 0$, the operators
$$
L_{n} \equiv {d\over d\tau}-{\kappa_{n}\over \tau}
\; \; \; \; ,
\eqno (2.17)
$$
$$
M_{n} \equiv {d\over d\tau}+{\kappa_{n}\over \tau}
\; \; \; \; ,
\eqno (2.18)
$$
the coupled eigenvalue equations take, in light of the
mode-by-mode expansion of the action integral,$^{4,9}$ the form
$$
L_{n}x=E{\widetilde x}
\; \; \; \; , \; \; \; \;
M_{n}{\widetilde x}=-Ex
\; \; \; \; ,
\eqno (2.19)
$$
$$
L_{n}y=E{\widetilde y}
\; \; \; \; , \; \; \; \;
M_{n}{\widetilde y}=-Ey
\; \; \; \; ,
\eqno (2.20)
$$
$$
L_{n}X=E{\widetilde X}
\; \; \; \; , \; \; \; \;
M_{n}{\widetilde X}=-EX
\; \; \; \; ,
\eqno (2.21)
$$
$$
L_{n}Y=E{\widetilde Y}
\; \; \; \; , \; \; \; \;
M_{n}{\widetilde Y}=-EY
\; \; \; \; ,
\eqno (2.22)
$$
where
$$
x \equiv m_{n1}^{(\beta)}
\; \; \; \; , \; \; \; \;
X \equiv m_{n2}^{(\beta)}
\; \; \; \; ,
\eqno (2.23)
$$
$$
{\widetilde x} \equiv {\widetilde m}_{n1}^{(\beta)}
\; \; \; \; , \; \; \; \;
{\widetilde X} \equiv {\widetilde m}_{n2}^{(\beta)}
\; \; \; \; ,
\eqno (2.24)
$$
$$
y \equiv r_{n1}^{(\mu)}
\; \; \; \; , \; \; \; \;
Y \equiv r_{n2}^{(\mu)}
\; \; \; \; ,
\eqno (2.25)
$$
$$
{\widetilde y} \equiv {\widetilde r}_{n1}^{(\mu)}
\; \; \; \; , \; \; \; \;
{\widetilde Y} \equiv {\widetilde r}_{n2}^{(\mu)}
\; \; \; \; .
\eqno (2.26)
$$
We now define $\forall n \geq 0$ the differential
operators
$$
P_{n} \equiv {d^{2}\over d\tau^{2}}+
\left[E^{2}-{\Bigr((n+3)^{2}-{1\over 4}\Bigr)\over
\tau^{2}}\right]
\; \; \; \; ,
\eqno (2.27)
$$
$$
Q_{n} \equiv {d^{2}\over d\tau^{2}}
+\left[E^{2}-{\Bigr((n+2)^{2}-{1\over 4}\Bigr)\over
\tau^{2}}\right]
\; \; \; \; .
\eqno (2.28)
$$
Eqs. (2.19)-(2.22) lead to the following second-order
equations, $\forall n\geq 0$:
$$
P_{n}{\widetilde x}=P_{n}{\widetilde X}
=P_{n}{\widetilde y}=P_{n}{\widetilde Y}=0
\; \; \; \; ,
\eqno (2.29)
$$
$$
Q_{n}y=Q_{n}Y=Q_{n}x=Q_{n}X=0
\; \; \; \; .
\eqno (2.30)
$$
The solutions of (2.29)-(2.30) regular at the origin are
$$
{\widetilde x}=C_{1}\sqrt{\tau}J_{n+3}(E\tau)
\; \; \; \; , \; \; \; \;
{\widetilde X}=C_{2}\sqrt{\tau}J_{n+3}(E\tau)
\; \; \; \; ,
\eqno (2.31)
$$
$$
x=C_{3}\sqrt{\tau}J_{n+2}(E\tau)
\; \; \; \; , \; \; \; \;
X=C_{4}\sqrt{\tau}J_{n+2}(E\tau)
\; \; \; \; ,
\eqno (2.32)
$$
$$
{\widetilde y}=C_{5}\sqrt{\tau}J_{n+3}(E\tau)
\; \; \; \; , \; \; \; \;
{\widetilde Y}=C_{6}\sqrt{\tau}J_{n+3}(E\tau)
\; \; \; \; ,
\eqno (2.33)
$$
$$
y=C_{7}\sqrt{\tau}J_{n+2}(E\tau)
\; \; \; \; , \; \; \; \;
Y=C_{8}\sqrt{\tau}J_{n+2}(E\tau)
\; \; \; \; .
\eqno (2.34)
$$
To find the condition obeyed by the eigenvalues $E$,
we now insert (2.31)-(2.34) into the boundary conditions
(2.13)-(2.16), taking into account also the first-order
system given by (2.19)-(2.22). This gives the eight
equations
$$
iC_{3}J_{n+2}(Ea)=\epsilon \; C_{2}J_{n+3}(Ea)
\; \; \; \; ,
\eqno (2.35)
$$
$$
iC_{4}J_{n+2}(Ea)=-\epsilon \; C_{1}J_{n+3}(Ea)
\; \; \; \; ,
\eqno (2.36)
$$
$$
iC_{5}J_{n+3}(Ea)=\epsilon \; C_{8}J_{n+2}(Ea)
\; \; \; \; ,
\eqno (2.37)
$$
$$
iC_{6}J_{n+3}(Ea)=-\epsilon \; C_{7}J_{n+2}(Ea)
\; \; \; \; ,
\eqno (2.38)
$$
$$
C_{1}=-{EC_{3}J_{n+2}(Ea)\over
\biggr[E{\dot J}_{n+3}(Ea)+(n+3){J_{n+3}(Ea)\over a}
\biggr]}
\; \; \; \; ,
\eqno (2.39)
$$
$$
C_{2}=-{EC_{4}J_{n+2}(Ea)\over
\biggr[E{\dot J}_{n+3}(Ea)+(n+3){J_{n+3}(Ea)\over a}
\biggr]}
\; \; \; \; ,
\eqno (2.40)
$$
$$
C_{7}={EC_{5}J_{n+3}(Ea)\over
\biggr[E{\dot J}_{n+2}(Ea)-(n+2){J_{n+2}(Ea)\over a}
\biggr]}
\; \; \; \; ,
\eqno (2.41)
$$
$$
C_{8}={EC_{6}J_{n+3}(Ea)\over
\biggr[E{\dot J}_{n+2}(Ea)-(n+2){J_{n+2}(Ea)\over a}
\biggr]}
\; \; \; \; .
\eqno (2.42)
$$
Interestingly, these give separate relations among the
constants $C_{1},C_{2},C_{3},C_{4}$ and among
$C_{5},C_{6},C_{7},C_{8}$ [3,9]. For example, eliminating
$C_{1},C_{2},C_{3},C_{4}$, using (2.35)-(2.36),
(2.39)-(2.40) and the useful identities$^{14}$
$$
Ea{\dot J}_{n+2}(Ea)-(n+2)J_{n+2}(Ea)=-EaJ_{n+3}(Ea)
\; \; \; \; ,
\eqno (2.43)
$$
$$
Ea{\dot J}_{n+3}(Ea)+(n+3)J_{n+3}(Ea)=EaJ_{n+2}(Ea)
\; \; \; \; ,
\eqno (2.44)
$$
one finds
$$
i\epsilon \; {J_{n+2}(Ea)\over J_{n+3}(Ea)}
=\epsilon^{2}{C_{2}\over C_{3}}=\epsilon^{2}
{C_{4}\over C_{1}}=i\epsilon^{3}
{J_{n+3}(Ea)\over J_{n+2}(Ea)}
\; \; \; \; ,
\eqno (2.45)
$$
which implies (since $\epsilon=\pm 1$)
$$
{\Bigr[J_{n+2}(E)\Bigr]}^{2}-{\Bigr[J_{n+3}(E)\Bigr]}^{2}=0
\; \; \; \; , \; \; \; \;
\forall n \geq 0
\; \; \; \; ,
\eqno (2.46)
$$
where we set $a=1$ for simplicity.
\vskip 1cm
\leftline {\bf 3. Physical-Degrees-of-Freedom Contribution to $\zeta(0)$}
\vskip 1cm
\noindent
The eigenvalue condition (2.46) is very similar to the formula
found in Refs. 3,9 for spin ${1\over 2}$, i.e.
${\Bigr[J_{n+1}(E)\Bigr]}^{2}-{\Bigr[J_{n+2}(E)\Bigr]}^{2}=0,
\forall n \geq 0$. Thus, the same technique
can be now applied to derive the PDF contribution to $\zeta(0)$ in the case
of gravitinos.
As we know from Refs. 4,9, the completely symmetric harmonics have
degeneracy $d(n)=(n+4)(n+1)$, $\forall n \geq 0$. This is the full degeneracy
in the case of local boundary conditions (2.1), since we need twice
as many modes to get the same number of eigenvalue conditions
as in the spectral case.$^{3-4,9}$ The $\zeta(0)$ calculation is now
performed using ideas first described in Ref. 15, and then used in
Refs. 3,9. Given the zeta-function at large $x$
$$
\zeta(s,x^{2}) \equiv \sum_{j=1}^{\infty}{\Bigr(\lambda_{j}
+x^{2}\Bigr)}^{-s}
\; \; \; \; ,
\eqno (3.1)
$$
where $\lambda_{j}=E^{2}$ are the squared eigenvalues of the
Dirac operator in our case,$^{3,9}$ one has in four dimensions
$$
\Gamma(3)\zeta(3,x^{2})=\int_{0}^{\infty}T^{2}e^{-x^{2}T}
G(T) \; dT \sim
\sum_{l=0}^{\infty}C_{l}\Gamma \Bigr(1+{l\over 2}\Bigr)
x^{-l-2}
\; \; \; \; ,
\eqno (3.2)
$$
where we have used the asymptotic expansion$^{9}$ of the heat
kernel for $T \rightarrow 0^{+}$
$$
G(T) \sim \sum_{l=0}^{\infty}C_{l}T^{{l\over 2}-2}
\; \; \; \; .
\eqno (3.3)
$$
On the other hand, defining $m \equiv n+3$, we find$^{3,9}$
$$ \eqalignno{
\Gamma(3)\zeta(3,x^{2})&=\sum_{m=3}^{\infty}(m+1)(m-2)
{\left({1\over 2x}{d\over dx}\right)}^{3}
\log \left[(ix)^{-2(m-1)}\left(J_{m-1}^{2}-J_{m}^{2}\right)(ix)\right] \cr
& \sim \sum_{m=0}^{\infty}\left(m^{2}-m\right)
{\left({1\over 2x}{d\over dx}\right)}^{3}
\left[\sum_{i=1}^{5}S_{i}(m,\alpha_{m}(x))\right]\cr
&+Z_{1}+Z_{2}+\sum_{n=5}^{\infty}q_{n}x^{-2-n}
\; \; \; \; ,
&(3.4)\cr}
$$
where$^{3,9}$
$$
\alpha_{m}(x) \equiv \sqrt{m^{2}+x^{2}}
\; \; \; \; ,
\eqno (3.5)
$$
$$
S_{1}(m,\alpha_{m}(x)) \equiv -\log(\pi)
+2\alpha_{m}
\; \; \; \; ,
\eqno (3.6)
$$
$$
S_{2}(m,\alpha_{m}(x)) \equiv -(2m-1)\log(m+\alpha_{m})
\; \; \; \; ,
\eqno (3.7)
$$
$$
S_{3}(m,\alpha_{m}(x)) \equiv \sum_{r=0}^{2}
k_{1r}m^{r}\alpha_{m}^{-r-1}
\; \; \; \; ,
\eqno (3.8)
$$
$$
S_{4}(m,\alpha_{m}(x)) \equiv \sum_{r=0}^{4}
k_{2r}m^{r}\alpha_{m}^{-r-2}
\; \; \; \; ,
\eqno (3.9)
$$
$$
S_{5}(m,\alpha_{m}(x)) \equiv \sum_{r=0}^{6}
k_{3r}m^{r}\alpha_{m}^{-r-3}
\; \; \; \; ,
\eqno (3.10)
$$
$$
Z_{1}\equiv -2\sum_{m=0}^{\infty}
{\left({1\over 2x}{d\over dx}\right)}^{3}
\left[\sum_{i=1}^{5}S_{i}(m,\alpha_{m}(x))\right]
=\sum_{i=1}^{5}X_{\infty}^{(i)}
\; \; \; \; ,
\eqno (3.11)
$$
$$
Z_{2}\equiv 2\sum_{m=0}^{1}
{\left({1\over 2x}{d\over dx}\right)}^{3}
\left[\sum_{i=1}^{5}S_{i}(m,\alpha_{m}(x))\right]
=\sum_{i=1}^{5}Y_{\infty}^{(i)} \; \; \; \; .
\eqno (3.12)
$$
One can thus obtain $\zeta(0)=C_{4}$ as half the coefficient
of $x^{-6}$ in the asymptotic expansion of the right-hand
side of (3.4), by comparison of (3.2) and (3.4), and bearing
in mind that$^{3,9}$
$$
k_{10}=-{1\over 4} \; \; \; \; , \; \; \; \;
k_{11}=0 \; \; \; \; , \; \; \; \;
k_{12}={1\over 12}
\; \; \; \; ,
\eqno (3.13)
$$
$$
k_{20}=0 \; \; , \; \;
k_{21}=-{1\over 8} \; \; , \; \;
k_{22}=k_{23}={1\over 8} \; \; , \; \;
k_{24}=-{1\over 8}
\; \; \; \; ,
\eqno (3.14)
$$
$$
k_{30}={5\over 192} \; \; , \; \;
k_{31}=-{1\over 8} \; \; , \; \;
k_{32}={9\over 320} \; \; , \; \;
k_{33}={1\over 2}
\; \; \; \; ,
\eqno (3.15a)
$$
$$
k_{34}=-{23\over 64} \; \; \; \; , \; \; \; \;
k_{35}=-{3\over 8} \; \; \; \; , \; \; \; \;
k_{36}={179\over 576} \; \; \; \; .
\eqno (3.15b)
$$
The PDF $\zeta(0)$ value for spin ${3\over 2}$ is thus given
by the spin-${1\over 2}$ value first found in Ref. 3 plus the
contributions of $Z_{1}$ and $Z_{2}$. For this purpose, we
also use the identities$^{3,9,15}$
$$
{\biggr({1\over 2x}{d\over dx}\biggr)}^{3}
\log{\biggr({1\over m+\alpha_{m}}\biggr)}
=(m+\alpha_{m})^{-3}\biggr[-\alpha_{m}^{-3}
-{9\over 8}m\alpha_{m}^{-4}-{3\over 8}m^{2}
\alpha_{m}^{-5}\biggr]
\; \; ,
\eqno (3.16)
$$
$$
(m+\alpha_{m})^{-3}={(\alpha_{m}-m)^{3}\over x^{6}}
\; \; \; \; .
\eqno (3.17)
$$
The insertion of (3.17) into (3.16) yields$^{3,9,15}$
$$
{\biggr({1\over 2x}{d\over dx}\biggr)}^{3}
\Bigr[-m\log(m+\alpha_{m})\Bigr]
=-mx^{-6}+m^{2}x^{-6}\alpha_{m}^{-1}
+{m^{2}\over 2}x^{-4}\alpha_{m}^{-3}
+{3\over 8}m^{2}x^{-2}\alpha_{m}^{-5}
\; \; .
\eqno (3.18)
$$
This further identity leads to divergences in the calculation,
but these are only {\it fictitious} in light of (3.16).
Such fictitious divergences are regularized dividing by
$\alpha_{m}^{2s}$, summing using the contour formulae$^{3,9,15}$
$$
\sum_{m=0}^{\infty}m^{2k}\alpha_{m}^{-2k-q}
={\Gamma \Bigr(k+{1\over 2}\Bigr)
\Gamma \Bigr({q\over 2}-{1\over 2}\Bigr) \over
2\Gamma \Bigr(k+{q\over 2}\Bigr)}
x^{1-q} \; \; , \; \;
\forall k=1,2,3,...
\eqno (3.19)
$$
$$
\sum_{m=0}^{\infty}m\alpha_{m}^{-1-q}
\sim {x^{1-q}\over \sqrt{\pi}}
\sum_{r=0}^{\infty}{2^{r}\over r!}B_{r}x^{-r}
{\Gamma \Bigr({r\over 2}+{1\over 2}\Bigr)
\Gamma \Bigr({q\over 2}-{1\over 2}+{r\over 2}\Bigr)
\over
2\Gamma \Bigr({1\over 2}+{q\over 2}\Bigr)}
\cos \Bigr({r\pi \over 2}\Bigr)
\; \; \; \; ,
\eqno (3.20)
$$
where $B_{r}$ are Bernoulli numbers,
and then taking the limit $s \rightarrow 0$.$^{3,9,15}$

Indeed, from (3.11) we find
$$
X_{\infty}^{(1)}=-{3\over 2}\sum_{m=0}^{\infty}
\alpha_{m}^{-5}
\; \; \; \; ,
\eqno (3.21)
$$
which does not contain $x^{-6}$ and hence does not contribute to $\zeta(0)$.
However, (3.18) and (3.7) imply
$$
X_{\infty}^{(2)}=4x^{-6}\beta_{1}-4x^{-6}\beta_{2}-2x^{-4}\beta_{3}
-{3\over 2}x^{-2}\beta_{4}-2x^{-6}\beta_{5}+2x^{-6}\beta_{6}+x^{-4}\beta_{7}
+{3\over 4}x^{-2}\beta_{8}
\; ,
\eqno (3.22)
$$
where
$$
\beta_{1} \equiv \sum_{m=0}^{\infty}m
\; \; \; \; ,
\eqno (3.23)
$$
$$
\beta_{2} \equiv \sum_{m=0}^{\infty}
m^{2}\alpha_{m}^{-1}
\; \; \; \; ,
\eqno (3.24)
$$
$$
\beta_{3} \equiv \sum_{m=0}^{\infty}m^{2}\alpha_{m}^{-3}
\; \; \; \; ,
\eqno (3.25)
$$
$$
\beta_{4} \equiv \sum_{m=0}^{\infty}m^{2}\alpha_{m}^{-5}
\; \; \; \; ,
\eqno (3.26)
$$
$$
\beta_{5}\equiv \lim_{s \to 0}\sum_{m=0}^{\infty}\alpha_{m}^{-2s}
\; \; \; \; ,
\eqno (3.27)
$$
$$
\beta_{6}\equiv \lim_{s \to 0}\sum_{m=0}^{\infty}
m\alpha_{m}^{-1-2s}
\; \; \; \; ,
\eqno (3.28)
$$
$$
\beta_{7} \equiv \sum_{m=0}^{\infty}m\alpha_{m}^{-3}
\; \; \; \; ,
\eqno (3.29)
$$
$$
\beta_{8} \equiv \sum_{m=0}^{\infty}m\alpha_{m}^{-5}
\; \; \; \; .
\eqno (3.30)
$$
Note that only $\beta_{1}$ and $\beta_{5}$ contribute to
$\zeta(0)$. This is proved using (3.19)-(3.20) and the
Euler-Maclaurin formula. According to this algorithm, if $f$
is a real- or complex-valued function defined on
$0 \leq t \leq \infty$, and if $f^{(2m)}(t)$ is absolutely
integrable on $(0,\infty)$ then, for $u=1,2,...$ $^{3,9,16}$
$$
\sum_{i=0}^{u}f(i)-\int_{0}^{u}f(x) \; dx
={1\over 2}\Bigr[f(0)+f(u)\Bigr]
+\sum_{s=1}^{m-1}{B_{2s}\over (2s)!}
\biggr[f^{(2s-1)}(u)-f^{(2s-1)}(0)\biggr]
+R_{m}(u)
\; ,
\eqno (3.31)
$$
where the remainder $R_{m}$ satisfies the inequality
$$
\mid R_{m}(u) \mid \leq \Bigr(2-2^{1-2m}\Bigr)
{\mid B_{2m} \mid \over (2m)!}
\int_{0}^{u}\mid f^{(2m)}(x)\mid \; dx
\; \; \; \; .
\eqno (3.32)
$$
The asymptotic expansion (3.20) implies that
$\beta_{1}$ gives the contribution
$$
\delta^{(a)}=2 \cos(\pi){\Gamma \left({3\over 2}\right)\over
\Gamma \left({1\over 2}\right)}B_{2}=-{1\over 6}
\; \; \; \; ,
\eqno (3.33)
$$
and the Euler-Maclaurin formula shows that $\beta_{5}$ contributes
$$
\delta^{(b)}=-{1\over 2}
\; \; \; \; .
\eqno (3.34)
$$

By virtue of (3.13), (3.8) and (3.11), we also find that
$$
X_{\infty}^{(3)}={15\over 4}k_{10}\sum_{m=0}^{\infty}
\alpha_{m}^{-7}
+{105\over 4}k_{12}\sum_{m=0}^{\infty}m^{2}
\alpha_{m}^{-9}
\; \; \; \; .
\eqno (3.35)
$$
Thus, using (3.19) and (3.13), we derive the following contribution to
$\zeta(0)$:
$$
\delta^{(c)}=(k_{10}+k_{12})=-{1\over 6}
\; \; \; \; .
\eqno (3.36)
$$

Finally, using (3.9)-(3.11) we obtain
$$
X_{\infty}^{(4)}={1\over 4}\sum_{r=0}^{4}k_{2r}(r+2)(r+4)(r+6)
\left[\sum_{m=0}^{\infty}m^{r}\alpha_{m}^{-r-8}\right]
\; \; \; \; ,
\eqno (3.37)
$$
$$
X_{\infty}^{(5)}={1\over 4}\sum_{r=0}^{6}k_{3r}(r+3)(r+5)(r+7)
\left[\sum_{m=0}^{\infty}m^{r}\alpha_{m}^{-r-9}\right]
\; \; \; \; ,
\eqno (3.38)
$$
and in light of (3.19)-(3.20) we derive that the asymptotic
behaviour of $X_{\infty}^{(4)}$ is ${\rm O}(x^{-7})$,
and the asymptotic form of $X_{\infty}^{(5)}$
is ${\rm O}(x^{-8})$.
Thus, they do not affect the $\zeta(0)$ value.

Moreover, the whole of $Z_{2}$ (cf (3.12)) does not
affect $\zeta(0)$. In fact one finds
$$
Y_{\infty}^{(1)}={3\over 2}x^{-5}\left[1+{\Bigr(1+x^{-2}\Bigr)}^
{-{5\over 2}}\right]
\; \; \; \; ,
\eqno (3.39)
$$
$$
Y_{\infty}^{(2)}=2x^{-7}{\Bigr(1+x^{-2}\Bigr)}^{-{1\over 2}}
+x^{-7}{\Bigr(1+x^{-2}\Bigr)}^{-{3\over 2}}+{3\over 4}x^{-7}
{\Bigr(1+x^{-2}\Bigr)}^{-{5\over 2}}
\; \; \; \; ,
\eqno (3.40)
$$
$$
Y_{\infty}^{(3)}=-{15\over 4}k_{10}x^{-7}
\left[1+{\Bigr(1+x^{-2}\Bigr)}^{-{7\over 2}}\right]
-{105\over 4}k_{12}x^{-9}{\Bigr(1+x^{-2}\Bigr)}^{-{9\over 2}}
\; \; \; \; ,
\eqno (3.41)
$$
$$
Y_{\infty}^{(4)}=-{1\over 4}\sum_{r=1}^{4}x^{-r-8}k_{2r}(r+2)(r+4)(r+6)
{\Bigr(1+x^{-2}\Bigr)}^{-{r\over 2}-4}
\; \; \; \; ,
\eqno (3.42)
$$
$$
Y_{\infty}^{(5)}=-{105\over 4}k_{30}x^{-9}-{1\over 4}
\sum_{r=0}^{6}x^{-r-9}k_{3r}(r+3)(r+5)(r+7)
{\Bigr(1+x^{-2}\Bigr)}^{-{r\over 2}-{9\over 2}}
\; \; \; \; ,
\eqno (3.43)
$$
and the reader can now easily see that
the formulae (3.39)-(3.43) do not
contain terms proportional to $x^{-6}$.

At the end, we have to consider more carefully the effect of
higher-order terms in the asymptotic expansion
of $\log \biggr[(ix)^{-2(m-1)}\Bigr(J_{m-1}^{2}-J_{m}^{2}
\Bigr)(ix)\biggr]$.
In light of Refs. 3,9 and of Eqs. (3.4)-(3.12) we study, $\forall n>3$
$$ \eqalignno{
{\widetilde H}_{\infty}^{n,A}&\equiv -{1\over 4}\sum_{p=1}^{l}h_{np}
\sum_{m=0}^{\infty}\Bigr[a_{np}\alpha_{m}^{p-n-6}
(m+\alpha_{m})^{-p}
+b_{np}\alpha_{m}^{p-n-5}(m+\alpha_{m})^{-p-1}\cr
&+c_{np}\alpha_{m}^{p-n-4}
(m+\alpha_{m})^{-p-2}+d_{np}\alpha_{m}^{p-n-3}
(m+\alpha_{m})^{-p-3}\Bigr]
\; \; \; \; ,
&(3.44)\cr}
$$
$$
{\widetilde H}_{\infty}^{n,B}\equiv {1\over 4}\sum_{r=0}^{2n}k_{nr}
(r+n)(r+n+2)(r+n+4)\sum_{m=0}^{\infty}m^{r}\alpha_{m}^{-r-n-6}
\; \; \; \; ,
\eqno (3.45)
$$
$$ \eqalignno{
{\widetilde H}_{\infty}^{n,C}&\equiv {1\over 4}\sum_{p=1}^{l}h_{np}
\sum_{m=0}^{1}\Bigr[a_{np}\alpha_{m}^{p-n-6}
(m+\alpha_{m})^{-p}
+b_{np}\alpha_{m}^{p-n-5}(m+\alpha_{m})^{-p-1}\cr
&+c_{np}\alpha_{m}^{p-n-4}(m+\alpha_{m})^{-p-2}
+d_{np}\alpha_{m}^{p-n-3}
(m+\alpha_{m})^{-p-3}\Bigr]
\; \; \; \; ,
&(3.46)\cr}
$$
$$
{\widetilde H}_{\infty}^{n,D}\equiv -{1\over 4}\sum_{r=0}^{2n}
k_{nr}(r+n)(r+n+2)(r+n+4)\sum_{m=0}^{1}m^{r}
\alpha_{m}^{-r-n-6}
\; \; \; \; ,
\eqno (3.47)
$$
where $a_{np}, b_{np}, c_{np}, d_{np}, h_{np}$ are constant coefficients.
In (3.44)-(3.47), $n$ should not
be confused with the integer appearing in
(2.46) and in the definition of $m$.
Again, the Euler-Maclaurin formula is very useful in studying
${\widetilde H}_{\infty}^{n,A}$. The equivalent of $f(0)$ in (3.31) gives a
contribution proportional to $x^{-n-6}$. Bernoulli numbers and derivatives
of odd order give a contribution proportional to $x^{-n-7}$ plus higher-order
terms. The conversion of (3.44) into an integral yields a term
proportional to $x^{-n-5}$, as it is evident studying the
integrals
$$
{\widetilde I}_{1}^{(np)} \equiv
\int_{0}^{\infty}
{\biggr(y+\sqrt{x^{2}+y^{2}}\biggr)}^{-p}
{\Bigr(x^{2}+y^{2}\Bigr)}^{{p\over 2}-{n\over 2}-3}
\; dy
\; \; \; \; ,
\eqno (3.48)
$$
$$
{\widetilde I}_{2}^{(np)} \equiv
\int_{0}^{\infty}
{\biggr(y+\sqrt{x^{2}+y^{2}}\biggr)}^{-p-1}
{\Bigr(x^{2}+y^{2}\Bigr)}^{{p\over 2}-{n\over 2}-{5\over 2}}
\; dy
\; \; \; \; ,
\eqno (3.49)
$$
$$
{\widetilde I}_{3}^{(np)} \equiv
\int_{0}^{\infty}
{\biggr(y+\sqrt{x^{2}+y^{2}}\biggr)}^{-p-2}
{\Bigr(x^{2}+y^{2}\Bigr)}^{{p\over 2}-{n\over 2}-2}
\; dy
\; \; \; \; ,
\eqno (3.50)
$$
$$
{\widetilde I}_{4}^{(np)} \equiv
\int_{0}^{\infty}
{\biggr(y+\sqrt{x^{2}+y^{2}}\biggr)}^{-p-3}
{\Bigr(x^{2}+y^{2}\Bigr)}^{{p\over 2}-{n\over 2}
-{3\over 2}}
\; dy
\; \; \; \; .
\eqno (3.51)
$$
The effect of ${\widetilde H}_{\infty}^{n,B}$ is derived by
using (3.19)-(3.20). When $r=0$ we have to consider $\sum_{m=0}^{\infty}
\alpha_{m}^{-n-6}$, which does not contain $x^{-6}$. When $r=2k>0$, (3.19)
leads to a contribution proportional to $x^{-n-5}$, and when $r=2k+1$,
(3.20) leads to a contribution proportional to $x^{-n-5}$ plus higher-order
terms. One also finds that
$$ \eqalignno{
{\widetilde H}_{\infty}^{n,C}&={x^{-n-6}\over 4}\sum_{p=1}^{l}h_{np}
\biggr[(a_{np}+b_{np}+c_{np}+d_{np})\cr
&+a_{np}{\Bigr(1+x^{-2}\Bigr)}^{{p\over 2}-{n\over 2}-3}
{\Bigr(x^{-1}+\sqrt{1+x^{-2}}\Bigr)}^{-p}\cr
&+b_{np}{\Bigr(1+x^{-2}\Bigr)}^{{p\over 2}-{n\over 2}-{5\over 2}}
{\Bigr(x^{-1}+\sqrt{1+x^{-2}}\Bigr)}^{-p-1}\cr
&+c_{np}{\Bigr(1+x^{-2}\Bigr)}^{{p\over 2}-{n\over 2}-2}
{\Bigr(x^{-1}+\sqrt{1+x^{-2}}\Bigr)}^{-p-2}\cr
&\left. +d_{np}{\Bigr(1+x^{-2}\Bigr)}^{{p\over 2}-{n\over 2}-{3\over 2}}
{\Bigr(x^{-1}+\sqrt{1+x^{-2}}\Bigr)}^{-p-3}\right]
\; \; \; \; ,
&(3.52)\cr}
$$
$$ \eqalignno{
{\widetilde H}_{\infty}^{n,D}&=-{1\over 4}k_{n0} \; n(n+2)(n+4)x^{-n-6}
\left[1+{\Bigr(1+x^{-2}\Bigr)}^{-{n\over 2}-3}\right]\cr
&-{1\over 4}\sum_{r=1}^{2n}k_{nr}(r+n)(r+n+2)(r+n+4)x^{-r-n-6}
{\Bigr(1+x^{-2}\Bigr)}^{-{r\over 2}-{n\over 2}-3}.
&(3.53)\cr}
$$
This is why ${\widetilde H}_{\infty}^{n,A}$, ${\widetilde H}_{\infty}^{n,B}$,
${\widetilde H}_{\infty}^{n,C}$ and ${\widetilde H}_{\infty}^{n,D}$ do not
contain terms proportional to $x^{-6}$, and hence do not contribute to
$\zeta(0)$.

To sum up, in light of (3.4),
(3.33)-(3.34), (3.36), (3.44)-(3.47), and using
the $\zeta(0)$ value obtained in Ref. 3, we find
$$
\zeta(0)={11\over 360}-{5\over 6}=-{289\over 360}
\; \; \; \; ,
\eqno (3.54)
$$
which is equal to the PDF value found in Ref. 4 when one sets to
zero on $S^3$ all untwiddled coefficients of $\psi_{i}^{A}$ and
${\widetilde \psi}_{i}^{A'}$. However, as shown in Ref. 10,
$\zeta(0)$ values depend on the boundary conditions if
Majorana fermions and gravitinos are massive.
\vskip 1cm
\leftline {\bf 4. Concluding Remarks}
\vskip 1cm
\noindent
The calculation appearing in our paper was not performed explicitly
in Refs. 5,10, and was only available
in Ref. 9. We have therefore tried
to present it in a self-contained way in this journal, to make it
accessible to a wider audience. Interestingly, if the gauge
constraints (1.2) and supersymmetry constraints are imposed
{\it before} quantization, the PDF value is found to be
$\zeta^{(PDF)}(0)=-{289\over 360}$. However,
Becchi-Rouet-Stora-Tyutin-invariant
quantization techniques might lead to different $\zeta(0)$ values.
This is indeed what happens in Ref. 2, where, studying the effect
of ghost fields and gauge degrees of freedom, the author finds
$\zeta_{3\over 2}(0)={197\over 180}$. In this case the difference with
respect to the PDF value (3.54) is substantial, at least because the
signs are opposite. However, one should bear in mind that the discrepancy
found in Ref. 3 for the spin-${1\over 2}$ result also affects the
spin-${3\over 2}$ calculation.
Moreover, it is also worth remarking that in Ref. 2 the gravitino
contribution to $\zeta(0)$ in simple supergravity makes the one-loop
amplitude even more divergent, when perturbative modes for the
three-metric are set to zero on $S^3$.
By contrast, within the PDF approach, the
gravitino contribution to $\zeta(0)$ in $N=1$ supergravity partially
cancels the contribution of the gravitational field in such a case.

Our result (3.54) may not only
add evidence in favour of different quantization techniques
for gauge fields being inequivalent, but remains of some value
if a mode-by-mode gauge-invariant $\zeta(0)$ calculation is
performed. In that case, the physical degrees of freedom decouple
from gauge and ghost modes, so that their contribution to
$\zeta(0)$ is again given by equation (3.54) if the boundary
conditions (2.1) are required. Unfortunately, already in the
simpler case of Euclidean Maxwell theory in four
dimensions, gauge modes are then found to obey a very complicated
set of coupled eigenvalue equations, and it is not yet clear how
to evaluate their contribution to the full $\zeta(0)$ value in a
mode-by-mode analysis.$^{9}$ If this last technical problem could be
solved, one would then obtain a very relevant check of $\zeta(0)$
values for gauge fields in the presence of boundaries
previously found in the literature. Of course, supergravity
multiplets cannot be studied at one-loop about flat Euclidean
four-space, since the existence of a cosmological constant
is incompatible with a flat background geometry.$^{9}$
However, we hope that the calculations in our paper
(see also Ref. 10) can be used as a first step towards a
mode-by-mode perturbative analysis in the presence of curved
backgrounds, at least in the limit of small boundary
three-geometry.$^{9,17}$

A further interesting question, arising from the work in
Refs. 9,18-19, is whether local boundary conditions involving
{\it field strengths} rather than potentials can be used
for spin ${3\over 2}$. It is not yet clear whether,
and eventually how, the corresponding one-loop calculation
might be performed.
\vskip 1cm
\leftline {\bf Acknowledgments}
\vskip 1cm
\noindent
I am much indebted to Peter D'Eath for enlightening conversations
and to Alexander Kamenshchik for correspondence.
\vskip 1cm
\leftline {\bf References}
\vskip 1cm
\item {1.}
H. C. Luckock and I. G. Moss, {\it Class. Quantum Grav.}
{\bf 6}, 1993 (1989).
\item {2.}
S. Poletti, {\it Phys. Lett.} {\bf B249}, 249 (1990).
\item {3.}
P. D. D'Eath and G. Esposito, {\it Phys. Rev.}
{\bf D43}, 3234 (1991).
\item {4.}
P. D. D'Eath and G. Esposito, {\it Phys. Rev.}
{\bf D44}, 1713 (1991).
\item {5.}
P. D. D'Eath and G. Esposito,
in {\it Proceedings of the
9th Italian Conference on General Relativity and
Gravitational Physics}, eds. R. Cianci {\it et al.}
(World Scientific, Singapore, 1991).
\item {6.}
A. O. Barvinsky, A. Y. Kamenshchik, I. P. Karmazin and
I. V. Mishakov, {\it Class. Quantum Grav.} {\bf 9}, L27 (1992).
\item {7.}
A. Y. Kamenshchik and I. V. Mishakov, {\it Int. J. Mod.
Phys.} {\bf A7}, 3713 (1992).
\item {8.}
A. O. Barvinsky, A. Y. Kamenshchik and I. P. Karmazin,
{\it Ann. Phys.} {\bf 219}, 201 (1992).
\item {9.}
G. Esposito, {\it Quantum Gravity, Quantum Cosmology
and Lorentzian Geometries}, Lecture Notes in Physics,
New Series m: Monographs, Volume m12
(Springer-Verlag, Berlin, 1992).
\item {10.}
A. Y. Kamenshchik and I. V. Mishakov, {\it Phys. Rev.}
{\bf D47}, 1380 (1993).
\item {11.}
M. J. Duff, in
{\it Supergravity '81}, eds. S. Ferrara and J. G. Taylor
(Cambridge University Press, Cambridge, 1982).
\item {12.}
P. D. D'Eath, {\it Nucl. Phys.} {\bf B269}, 665 (1986).
\item {13.}
P. D. D'Eath and J. J. Halliwell, {\it Phys. Rev.}
{\bf D35}, 1100 (1987).
\item {14.}
I. S. Gradshteyn and I. M. Ryzhik, {\it Table of Integrals,
Series and Products} (Academic Press, New York, 1965).
\item {15.}
I. G. Moss, {\it Class. Quantum Grav.} {\bf 6}, 759 (1989).
\item {16.}
R. Wong, {\it Asymptotic Approximations of Integrals}
(Academic Press, New York, 1989).
\item {17.}
K. Schleich, {\it Phys. Rev.} {\bf D32}, 1889 (1985).
\item {18.}
P. Breitenlohner and D. Z. Freedman, {\it Ann. Phys.}
{\bf 144}, 249 (1982).
\item {19.}
S. W. Hawking, {\it Phys. Lett.} {\bf B126}, 175 (1983).
\bye